\documentclass[twocolumn,showpacs,amsmath,amssymb]{revtex4}
\usepackage{graphicx}

\begin{document}
\title{Observation of the Faraday effect via beam deflection in a longitudinal magnetic field}
\author{Ambarish Ghosh}
\affiliation{The Rowland Institute at Harvard, Harvard University,
Cambridge, Massachusetts 02142}
\author{Peer Fischer}
\affiliation{The Rowland Institute at Harvard, Harvard University,
Cambridge, Massachusetts 02142}

\begin{abstract}
We report the observation of the magnetic field induced circular
differential deflection of light at the interface of a Faraday
medium. The difference in the angles of refraction or reflection
between the two circular polarization components is a function of
the magnetic field strength and the Verdet constant. The reported
phenomena permit the observation of the Faraday effect not {\it
via} polarization rotation in transmission, but {\it via} changes
in the propagation direction in refraction or in reflection.  An
unpolarized light beam is predicted to split into its two circular
polarization components. The light deflection arises within a few
wavelengths at the interface and is therefore independent of
pathlength.
\end{abstract}

\maketitle

Well established magneto-optical phenomena are the Faraday,
Cotton-Mouton-Voigt and the magneto-optical Kerr (MOKE) effects,
as well as magnetic circular dichroism. These are described by
changes in the azimuth (optical rotation) or the ellipticity of an
electromagnetic wave.  Apart from changes in these Stokes
parameters, a magnetic field may also influence the propagation
direction of light.

The deflection of a light beam in isotropic media subject to a
homogenous \cite{inhomogenous} {\em transverse} magnetic field has
been reported by Rikken and Tiggelen who observed the deflection
in scattering \cite{rikken96} as well as in transmission
\cite{rikken97}, and by Blasberg and Suter \cite{suter} who showed
that angular momentum conservation causes a small lateral
displacement near resonance in an atomic vapor. Transverse
magnetic field induced refraction at the cesium vapor/glass
interface has been reported by Schlesser and Weis
\cite{schlesser}. However, their observed deflection appears to be
nonlinear in the strength of the magnetic field and nonlinear in
the light intensity \cite{schlesser}. A complete explanation for
this effect has not yet been provided \cite{schlesser,rikken97}.

Here, we show that it is also possible to observe magneto-optical
deflection at an interface in the presence of a {\em longitudinal}
magnetic field. In particular, we show that longitudinal magnetic
field induced refraction and reflection at an interface gives rise
to a circular differential beam deflection, and that this is an
alternate means to determine Verdet constants.

A magnetic field renders any medium (isotropic or oriented)
optically active. In particular, any isotropic medium becomes
uniaxial in the presence of a magnetic field, and its refractive
indices for right- $(+)$ and left- $(-)$ circularly polarized
light are unequal, such that the plane of polarization of a
linearly polarized electromagnetic wave rotates as the wave
propagates along the direction of the field. The Faraday rotation
in radians developed by an electromagnetic wave at the wavelength
$\lambda$ traversing a distance $l$ is given by:
\begin{equation}
\alpha  = V \, B \, l \; = \frac{\pi \, l}{\lambda} \left( n^{(-)}
- n^{(+)} \right) \;
\end{equation}
where $V$ is the frequency-dependent Verdet constant, and $B$ is
the magnetic field strength. Magneto-optical activity should,
however, not only manifest itself through Faraday rotation in
transmission, but should also be observable as a deflection of the
light beam in reflection \cite{silvermanamjp} and in refraction
 at an interface. We have recently reported the
observation of related phenomena in the case of natural optical
activity (chirality) \cite{splitting}.

\begin{figure}
\includegraphics[scale=0.8]{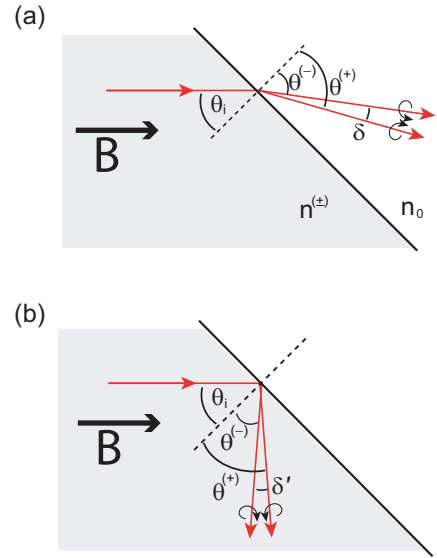}
\caption{a) Double refraction at the boundary of a Faraday medium
(shaded) and a medium with negligible Verdet constant
characterized by the scalar refractive index $n_0$. The direction
of the transmitted beam is different for left- and
right-circularly polarized waves. An unpolarized or a linearly
polarized beam will thus split into two. b) Reflection inside a
Faraday medium. The incident beam is subject to circular
birefringence $n^{(\pm)}$ such that the left- and right- circular
components have different angles of reflection.}
\end{figure}

If one considers the refraction of an electromagnetic wave at a
boundary formed by a Faraday medium and a medium with a negligible
Verdet constant, as shown in Fig. 1a, then the left- and the
right-circularly polarized waves or wave-components must
independently obey Snell's law. Circularly polarized waves that
propagate along the optical axis in the Faraday medium, will thus
refract in the medium characterized by the polarization
independent refractive index $n_0$ with angles of refraction
$\theta^{(-)}$ and $\theta^{(+)}$, depending on whether they are,
respectively, left- or right-circularly polarized. Similarly, if
an unpolarized or linearly polarized wave is incident from the
Faraday medium with angle of incidence $\theta_i$, then it will
split into two beams, one left- and the other right-circularly
polarized. The angular divergence $\delta= \theta^{(+)}-
\theta^{(-)}$ between the two refracted circular polarization
components in Fig. 1a is
\begin{equation}
\label{refraction} \delta \approx \frac{\left(
n^{(+)}-n^{(-)}\right)}{n_0} \, \frac{\sin\theta_i}{\cos\theta} \;
,
\end{equation}
where $\theta$ is the average of the two angles of refraction. It
follows that $B$-field induced deflection of light at an interface
can be used to determine circular birefringences, and hence Verdet
constants, or in the case of a known Verdet constant the strength
of an applied magnetic field:
\begin{equation}
\label{refraction} \delta \approx  - \frac{\lambda \;\,
\sin\theta_i}{\pi \, n_0 \, \cos\theta} \, V \, B \; .
\end{equation}
It is interesting to note that unlike Faraday rotation (Eq. (1)),
which is a function of the light path through the medium,
magneto-optical double refraction arises within a few wavelengths
at the interface. This could for instance be of use in the study
of ultrathin transparent samples \cite{splitting}.

One may also consider the components of an electromagnetic wave
that reflect inside the Faraday medium. Because a circularly
polarized wave reverses its circularity upon reflection, the
incident and the reflected waves are necessarily associated with
different refractive indices. Hence, in an optically active medium
the angle of reflection of a circularly polarized wave will in
general not equal the angle of incidence
\cite{silvermanamjp,splitting}. An unpolarized or linearly
polarized wave can therefore split into its two circularly
polarized components upon reflection \cite{splitting}. The
theoretical description of magnetic double reflection is
complicated by the fact that the reflected wave no longer
propagates along the optic axis of the system. The reflected beam
is thus potentially subject to circular birefringence (Faraday
effect) as well as the birefringence due to a transverse magnetic
field \cite{rikken97}. To simplify the discussion we will consider
a reflected beam that propagates in a direction perpendicular to
the magnetic field as shown in Fig 1b, such that it experiences no
birefringence due to the longitudinal component of the magnetic
field. Furthermore, we neglect any transverse $B$-field induced
birefringence and assume that the reflected waves are only subject
to an average refractive index $n=(n^{(-)} + n^{(+)})/2$. The
angular divergence $\delta'$ is then given by
\begin{equation}
\label{reflection} \delta' \approx - \frac{\lambda \,
\tan\theta_i}{\pi \, n}
 \, V \, B \; .
\end{equation}
Depending on the Fresnel reflection coefficients for a given
interface, the circular components may not fully reverse their
circularity upon reflection and become elliptically polarized.
This can be accounted for by including the appropriate Fresnel
coefficients.

\begin{figure}
\includegraphics[scale=0.18]{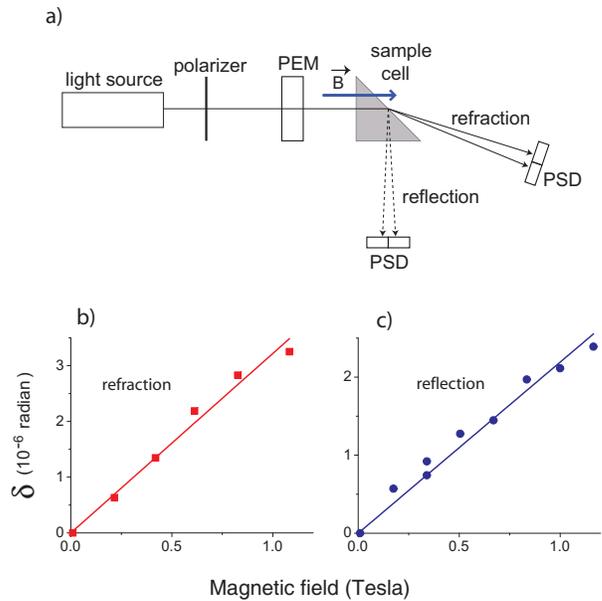}
\caption{a) Schematic of the experimental arrangement. Light
passes through a photoelastic modulator (PEM) where its
polarization is switched between left- and right-circular at
$\sim$50 kHz before refracting in a glass prism (solid line), or
reflecting inside a liquid cuvette (dotted line). The prism and
the prismatic cuvette are respectively mounted inside an
electromagnet (not shown). The position of the beam is recorded
with a position sensitive detector (PSD). b) Refraction data for
SF11 glass at 532 nm. c) Reflection data for pure CS$_2$ at
$\sim$473 nm.}
\end{figure}

We have observed the magnetic field induced double refraction and
reflection phenomena in an experimental arrangement schematically
depicted in Fig 2a. In the refraction measurements light from a
532 nm diode laser was modulated between left- and right-circular
polarized with a Hinds photoelastic modulator (PEM) at $\sim$ 50
kHz and then passed through a transparent glass (SF11) prism
mounted between the pole pieces of an electromagnet (Walker HV7).
The position of the laser beam was recorded with a position
sensitive diode and a lockin amplifer. The difference in the
angles of refraction for the two circular polarization components
is shown in Fig. 2b as a function of the applied magnetic field
strength. From a linear fit to the data and using Eqn. (3) a
Verdet constant of 29.37 $\pm$ 0.96 T rad$^{-1}$ m$^{-1}$ is
obtained. This is in good agreement with the tabulated values of
the Verdet constant for SF11 Schott glass \cite{Schott}, from
which we extrapolate a Verdet constant of 30.44 T rad$^{-1}$
m$^{-1}$. Each data point in Fig. 2b is measured with an
uncertainty that is approximately the size of the symbol. However,
the photoelastic modulator itself gives rise to an angular
deviation that fluctuates on the time scale of the measurement and
causes the data points to deviate from a straight line. The
goodness of the straight-line fit is thus a more appropriate
measure of the experimental error.
\\ The circular differential reflection in a longitudinal magnetic field was observed in a
right-angle prismatic cuvette filled with carbon disulfide ($\rm
CS_2$). A mirror was mounted parallel to the hypotenuse on the
inside of the liquid cuvette. The polarization modulated beam from
a $\sim$473 nm diode pumped solid state laser travelled along the
direction of the magnetic field in the liquid and upon reflection
exited the liquid perpendicular to the magnetic field and normal
to the window of the cuvette. From the detected angular divergence
shown in Fig. 2c we deduce a Verdet constant of 23.7 $\pm$ 0.8 T
rad$^{-1}$ m$^{-1}$. We suspect that the small difference with the
reported constant of 0.0694 min G$^{-1}$ cm$^{-1}$ (20.2 T
rad$^{-1}$ m$^{-1}$) at 476.5 nm \cite{verdetliquids} is due to
ellipticity in the reflected beam, which has not been accounted
for. We stress that even though the experimental geometry is
chosen such that the medium is (approximately) uniaxial, the
effects described here will in general be exhibited by any wave
that refracts or reflects at the interface of a Faraday medium.
Similar effects are expected to arise in diffraction
\cite{diffraction}.

In summary, we have shown that the Faraday effect can be observed
via double refraction or reflection at an interface in the
presence of a longitudinal magnetic field. We have demonstrated
that the difference in the propagation directions of the two
refracted (or reflected) circular polarization components is an
alternative means to determine Verdet constants or magnetic field
strengths.

The effects reported here distinguish themselves in a number of
ways from magneto-optical measurements reported hitherto. The
magnetic double refraction and reflection phenomena may be
observed both with polarized or with unpolarized electromagnetic
waves. Further, the effects arise within a few wavelengths at the
boundary, and may thus find application in the study of ultrathin
samples. Finally, the phenomena do not suffer from the n-$\pi$
ambiguity, which can plague Faraday rotation measurements in large
magnetic fields \cite{krushelnick}, or in space astronomy. This
prompts us to ask whether the deflection phenomena of this paper
also manifest themselves in astrophysical observations.

\end{document}